\documentclass[prl,twocolumn,draft,amsmath,showpacs]{revtex4}
\usepackage{graphics}

\begin{document}

\input epsf

\title{Spin thermal conductivity of the Haldane chain compound Y$_2$BaNiO$_5$}
\author{K. Kordonis,$^1$ A.V. Sologubenko,$^1$ T. Lorenz,$^1$ S.-W. Cheong,$^2$ and A. Freimuth$^1$}
\affiliation{$^1$II. Physikalisches Institut, Universit\"{a}t zu K\"{o}ln, 50937 K\"{o}ln, Germany}
\affiliation{$^2$Department of Physics and Astronomy, Rutgers University, Piscataway, New Jersey 08854}
\begin{abstract}
We have measured the thermal conductivity of the spin $S=1$ chain
compound Y$_2$BaNiO$_5$. Analyzing the anisotropy of the thermal
transport allows to identify a definite spin-mediated thermal
conductivity $\kappa_s$ along the chain direction. The calculated
spin-related energy diffusion constant $D_E(T)$ shows a broad
peak around 120~K. Close to room temperature,  $D_E(T)$
approaches the theoretically predicted high-temperature value,
while scattering of spin excitations by magnetic impurities seems
to be the major limiting factor of $\kappa_s$ at low temperature.
\end{abstract}
\pacs{
66.70.+f, % 6.70.+f Nonelectronic thermal conduction and heat-pulse propagation in solids
75.40.Gb % Dynamic properties (dynamic susceptibility, spin waves, spin diffusion, dynamic scaling, etc.)
75.47.-m %  Magnetotransport phenomena; materials for magnetotransport (for spintronics, see 85.75.-d; see also 72.15.Gd, 73.50.Jt, 73.43.Qt, and 72.25.-b in transport phenomena)
}
%\date{\today}
\maketitle

A broad class of low-dimensional quantum spin systems is expected
to exhibit anomalous transport properties due to the integrability
of the corresponding model Hamiltonians
\cite{Castella95,Saito96,Zotos96}. A particularly well studied
example is the one-dimensional Heisenberg antiferromagnetic (1D
HAF) $S=1/2$ XXZ model, for which several theoretical works
find finite Drude weights in the frequency spectra of the thermal
$\kappa(\omega)$ and the spin conductivity $\sigma(\omega)$ (see the recent review article \cite{Zotos05_Rev} and the references therein).
This, in turn, results in ballistic, dissipationless thermal and
spin conductivities.  These predictions are supported by
experimental observations of enhanced energy diffusion constants
$D_E$ and spin diffusion constants $D_S$ in several physical
realizations of the $S=1/2$ 1D HAF
model \cite{Takigawa96_213,Thurber01,Sologubenko01,Sologubenko03_Uni}.
In contrast,
in $S=1$ HAF chains the first excited state is separated
from the singlet ground state by an energy gap $\Delta$ and the
system is not integrable \cite{Haldane83}. This excludes infinite
conductivities, but anomalous diffusion constants may still be
expected \cite{Karadamoglou04}.
Experimental results on the thermal transport in $S=1$ chain
compounds are scarce. The existence of spin-mediated energy
transport in the $S=1$ chain compound AgVP$_2$S$_6$ has been
reported \cite{Sologubenko03_AVPS},  but the sample quality did
not allow for unambiguously resolving whether the cause of the
relatively low values of the spin-mediated thermal conductivity
is of intrinsic origin or is caused by defects.

In this Letter we report a study of the anisotropic thermal
conductivity of large high-quality single crystals of
Y$_2$BaNiO$_5$, which represents a nearly ideal realization of
the spin $S=1$ HAF chain. The basic building blocks
are chains of NiO$_6$ octahedra with Ni$^{2+}$ ions ($S=1$)
running along the $a$ axis of the orthorhombic structure with
lattice parameters $a=3.76$~\AA, $b=5.76$~\AA, and $c=11.33$~\AA
(space group Immm). The 1D character of the magnetic properties
is well established by susceptibility and inelastic neutron
scattering measurements yielding $J\approx 250 - 280$~K,
$\Delta\approx100$~K, and $|J_{\bot}/J|\leq10^{-4}$ for the
in-chain superexchange coupling, the Haldane gap, and the ratio
between out-of-chain and in-chain coupling, respectively
\cite{darriet93,batlogg94,yokoo95,sakaguchi96}. Our thermal
conductivity data provide clear evidence for a sizeable
spin-mediated energy transport along the chain direction, which
is absent along both perpendicular directions. In agreement with
theoretical expectations, the derived energy diffusion constant
shows a qualitatively different temperature dependence as those
found for various $S=1/2$ chain compounds. Our data also suggest
that magnetic impurities provide the dominating scattering
mechanism for magnons at low temperatures.

The single crystals of Y$_2$BaNiO$_5$ were grown by a self-flux
method \cite{Sulewski94}. Out of one crystal, a bar-shaped sample
was cut with dimensions $0.89 \times 0.79 \times 0.77$~mm$^3$
along the principal axes $a$, $b$, and $c$, respectively. For the
$\kappa(T)$ measurements we used a standard steady-state method \cite{Hofmann03}.
The thermal gradient of $\Delta T\simeq 0.1$~K has been
determined using a differential Chromel-Au+0.07\%Fe thermocouple.
Unaccounted heat losses due to the connecting leads and the sample
surface radiation are estimated to be negligible below about
200~K although at higher temperatures they can lead to somewhat
overestimated values of $\kappa$. For each orientation a few
$\kappa(T)$ runs have been made with removing and re-attaching
the thermal contacts to the sample. Different runs gave
qualitatively similar temperature dependencies, but different
absolute values varying up to $\pm$10\% for each orientation. This
mainly arises from the uncertainty in determining the exact
distance of the thermocouple junctions. Up to 14~T, we did not
find any magnetic-field dependence of the thermal conductivity.

%<<<<<<<<<<<<<<<<<<<<<<<< FIGURE 2 >>>>>>>>>>>>>>>>>>>>>>>>>
\begin{figure}[t]
 \begin{center}
  \leavevmode
  \epsfxsize=0.8\columnwidth \epsfbox {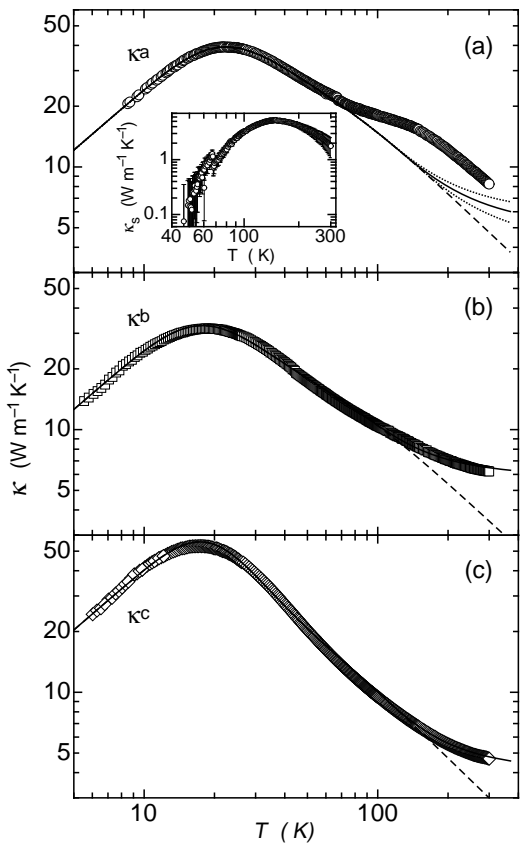}
   \caption{
   Thermal conductivity of Y$_2$BaNiO$_5$ along ($a$ axis)
   and perpendicular ($b$ and $c$ axis) to the spin chains.
   The solid (dashed) lines represent calculations of the
   phonon contributions with (without)
   taking into account a minimum phonon mean free path.
   The inset shows the spin contribution
   to the thermal conductivity along the chain direction (see text).
  }
\label{LamFit}
\end{center}
\end{figure}
%<<<<<<<<<<<<<<<<<<<<<<<< figure 2 >>>>>>>>>>>>>>>>>>>>>>>>>

In Fig.\ref{LamFit} we present typical results  for
$\kappa^{i}(T)$ ($i = a , b, c$) of Y$_2$BaNiO$_5$. Near 20~K,
$\kappa^{i}(T)$ for all three crystal directions exhibits a
low-temperature peak, which is typical for insulators where
thermal transport is of phononic origin. The peak signifies a
gradual change of the dominant phonon scattering mechanism from
sample boundaries and extended defects at low temperatures to
phonon-phonon scattering at high temperatures \cite{BermanBook}.
Another feature which is observed only for the $a$ direction is a
shoulder-like feature between about 80 and 140~K. Such an
anisotropy has been observed in various low-dimensional spin
systems such as 2D $S=1/2$ materials \cite{Nakamura91,Sales02,
Hofmann03,Hess03,Yan03,Jin03,Berggold06}, $S=1/2$ ladder
compounds \cite{Sologubenko00_lad,Hess01,Kudo01}, and $S=1/2$
chain compounds \cite{Ando98,Salce98,Sologubenko01,Ribeiro05}.
The conventional explanation is the existence of an extra
contribution to the thermal conductivity, beside the phonon
conductivity ($\kappa_{\rm ph}$), coming from spin excitations
($\kappa_s$), such that along the direction(s) of strong magnetic
exchange $\kappa^\parallel=\kappa_{\rm ph} + \kappa_s$. Because
of  weak interactions perpendicular to the chains
or planes, $\kappa_s$ is negligible in these directions, that is,
$\kappa^\perp \approx \kappa_{\rm ph}$. However, one has to be
careful in associating any high-temperature anomaly in
$\kappa^{\parallel}(T)$ with the spin transport, because
similarly looking features can also appear in a purely phononic thermal
conductivity as a result of scattering via spin-lattice
interaction in certain temperature regions \cite{Hofmann01}, as
well as being caused by additional thermal transport via optical
phonon modes \cite{Hess04}. There are no obvious reasons,
however, for these purely phononic mechanisms to show up only
along the spin-chain direction. Thus, the most likely origin of
the high-temperature feature in $\kappa^a(T)$ of Y$_2$BaNiO$_5$
is the spin contribution $\kappa_s$.

In order to analyze $\kappa_s(T)$ one has to subtract the phonon
contribution $\kappa^a_{\rm ph}(T)$ from the total measured
$\kappa^a(T)$ along the $a$ direction. Assuming that the same
scattering mechanisms influence $\kappa^i_{\rm ph}$ along all
directions, we first analyzed $\kappa^b(T)$ and $\kappa^c(T)$,
assumed to be purely phononic, in order to establish the main
phonon scattering mechanisms. Then we applied the same analysis
to $\kappa^a(T)$ at low temperatures where the spin contribution
$\kappa_s(T)$ is expected to be negligibly small due to the spin
gap, and extrapolated $\kappa^a_{\rm ph}(T)$ to higher
temperatures. For this analysis we employed the standard model of
phonon thermal conductivity \cite{BermanBook}:
\begin{equation}\label{eLambda}
    \kappa_{\rm ph} =  \frac{ k_{\rm B}}{ 2 {\pi}^{2} v } \left(
    \frac{k_{\rm     B}}{\hbar} \right) ^{3} T^{3}
    \int\limits_{0}^{\Theta_{D}/T}
    \frac{x^{4}e^{x}}{(e^{x}-1)^{2}}\tau (\omega,T) dx,
\end{equation}
where $x=\hbar\omega/k_{\rm B} T$, $\omega$ is the frequency of a
phonon, $\tau(\omega,T)$ is the relaxation time, $\Theta_{D}$ is
the Debye temperature,   $ v=\Theta _D\left( {{{k_B}
\mathord{\left/ {\vphantom {{k_B} \hbar }} \right.
\kern-\nulldelimiterspace} \hbar }} \right)(6\pi ^2n)^{-1/3}$ is
the mean sound velocity, and $n$ is the number density of atoms.
Due to the low $\kappa(T)\simeq 1$~W~m$^{-1}$~K$^{-1}$ at high
temperatures, which is comparable to $\kappa$ values of amorphous
materials \cite{Graebner86},  we took into account a lower limit
for the phonon mean free path $\ell_{\rm min}$. 
The underlying idea is that the phonon mean free path cannot be smaller than roughly the size of the unit cell. 
Such an $\ell_{\rm min}$ is
also observed in crystalline materials at high temperatures
\cite{Cahill92}. Thus, the equation for $\tau$ was put in the
form $ \tau(\omega,T) = \max\{\tau_{\Sigma}(\omega,T), \ell_{\rm
min}/v\}$. We now assume that all phonon scattering mechanisms
act independently, such that $ \tau_{\Sigma}^{-1} = \sum
\tau_{i}^{-1}$, where each  term  $\tau_{i}^{-1}$ corresponds to
an individual scattering mechanism. We found that two scattering
mechanisms, namely phonon scattering by extended two-dimensional
defects and phonon-phonon scattering, in the form
\begin{equation}\label{eTauR}
 \tau_{\Sigma}^{-1} = A \omega^2 + B \omega^2 T \exp\left( -\Theta_{D}/bT \right ),
\end{equation}
are sufficient to approximate $\kappa^b(T)$ and $\kappa^c(T)$. We
used $\Theta_{D}=390$~K estimated from the low-temperature
specific heat \cite{Ramirez94}. The parameters $A$, $B$, $b$, and
$\ell_{\rm min}$ were treated as freely adjustable parameters;
the fit values are presented in Table~\ref{t_1}.
%======================= TABLE 1 ==============================
\begin{table}[tb]
%\vskip-\lastskip
\caption{Parameters of the fits of $\kappa^i(T)$ to
Eq.~(\ref{eLambda}).} \label{t_1}
\begin{tabular}{lccc}
\colrule \multicolumn{1}{c}{Parameter} & \multicolumn{1}{c}{$a$
axis} & \multicolumn{1}{c}{$b$ axis} &
\multicolumn{1}{c}{$c$ axis} \\
\colrule
$A$ ($10^{-17}$ s$^{1}$)        & 4.0   &   3.8     &  2.4 \\
$B$ ($10^{-18}$ s K$^{-1}$)     & 9.1   &   12.0    &  14.9 \\
$b$                                 & 6.7   &   8.1     &  6.1  \\
$\ell_{\rm min}$  (\AA)         &           &   18.3    &  12.8  \\
Temperature region (K)    &  5--40  &   5--300 &   5--300   \\
\colrule
\end{tabular}
\end{table}
%-------------------------- table 1-------------------------------
The fits of $\kappa^i_{\rm ph}(T)$ for $i=b$ and $c$ are shown in
Fig.~\ref{LamFit}~(b,c) as solid lines.  The dashed lines are
obtained for the same values of $A$, $B$, and $b$, but setting
$\ell_{\rm min} = 0$. This demonstrates that $\ell_{\rm min}$
does not influence $\kappa^i_{\rm ph}(T)$ below about 100~K.
Hence, only the parameters $A$, $B$, $b$ can be adjusted by the
fit of $\kappa^a_{\rm ph}(T)$ in the temperature range $T < 50
{\rm ~K}$. We do not attribute 
much significance to the absolute values of the fit parameters for two reasons: 
(i) the Debye model is a crude estimate of the phonon spectrum of a real material,
and (ii) the fit is only used to estimate the phonon background
contribution, which has to be subtracted in order to separate the
spin-mediated thermal conductivity.

The spin thermal conductivity along the chains
$\kappa_s(T)=\kappa^a-\kappa^a_{\rm ph}$ is plotted in the inset
of Fig.~\ref{LamFit}~(a). For $\kappa^a_{\rm ph}(T)$ we used the
solid line of Fig.~\ref{LamFit}~(a) which is calculated for the
fit parameters of Table~\ref{t_1} and $\ell_{\rm min}=15.5 {\rm
~\AA}$, which is the average of the $\ell_{\rm min}$ values for
the $b$ and $c$ axes. The dotted lines are obtained for
$\kappa^a_{\rm ph}(T)$ if the individual values of $\ell_{\rm
min}$ for $b$ and $c$ are used. One can see that the errors of $\kappa^a_{\rm ph}$ due
to the uncertainty of $\ell_{\rm min}$ are negligible below about
140~K and reach about $\pm 10\%$ at 300~K. 
Below about 70~K, $\kappa_s$ is
not well resolved  because the difference between the total measured $\kappa^a$ and the estimated phonon contribution $\kappa^a_{\rm ph}$
becomes of the order of relative experimental errors, 
therefore we will consider only
the $\kappa_s$ data above this temperature. The maximum of $\kappa_s(T)$
amounts to about 5~W~K$^{-1}$~m$^{-1}$. That is about five times
higher than that observed for the only other Haldane system
AgVP$_2$S$_6$ ($J\simeq780$ K; $\Delta\simeq240$ K) for which
thermal conductivity has been investigated so far
\cite{Sologubenko03_AVPS}. From our $\kappa_s(T)$ data, we
calculated  the spin-related energy diffusion constant $D_{E}(T)
= \kappa_{s}(T) / (C_{s}(T) a^{2})$, where $C_{s}(T)$ is the
specific heat of the spin chain and $a$ the lattice constant
along the chain. For $C_{s}(T)$, the numerical results for the
ideal HAF $S=1$ chain with $J/k_B=280 {\rm ~K}$, calculated
within the quantum nonlinear $\sigma$-model (NL$\sigma$M)
\cite{Jolicoeur94Priv}, were used. 

%<<<<<<<<<<<<<<<<<<<<<<<< FIGURE 2 >>>>>>>>>>>>>>>>>>>>>>>>>
\begin{figure}[t]
 \begin{center}
  \leavevmode
  \epsfxsize=0.8\columnwidth \epsfbox {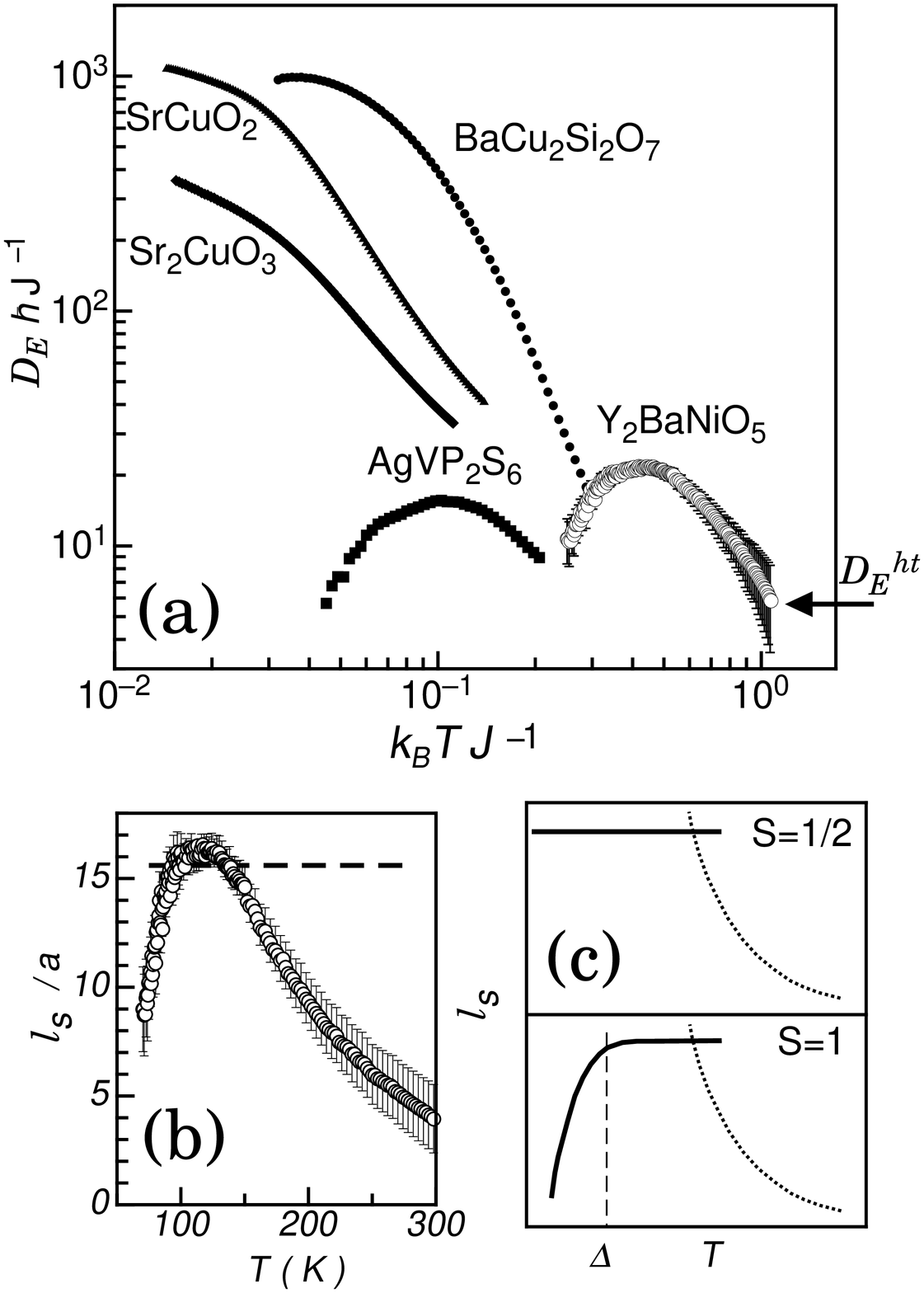}
   \caption{
  (a) The energy diffusion constant $D_{E}(T)$, calculated from
  the thermal conductivity data of Y$_2$BaNiO$_5$. For comparison,
 we also show $D_{E}(T)$ of the $S=1$ chain compound AgVP$_2$S$_6$
 and of the  $S=1/2$ chain compounds
 BaCu$_2$Si$_2$O$_7$  and SrCuO$_2$ (data from
 Refs.~\onlinecite{Sologubenko03_AVPS,Sologubenko03_Uni,Sologubenko01}).
 The arrow corresponds to the
 high-temperature limit $D^{ht}_E$ calculated in \cite{Karadamoglou04}.
 (b) The normalized mean
 free path of the spin excitations calculated from the thermal
conductivity data. The lower limit $\ell_{\rm min}^{\rm segment}$
of the average chain segment length estimated from the susceptibility data
is shown by the dashed line. 
(c)  Illustration of the expected differences in defect scattering in the $S=1/2$ and $S=1$ chains. The solid lines correspond to the mean free path for spin excitations  when defect scattering is dominant. The dotted lines corresponds to a temperature-dependent scattering which is expected to become dominant at high temperatures, such as interaction between spin excitations or interactions with phonons.  
  }
\label{DeJh}
\end{center}
\end{figure}
%<<<<<<<<<<<<<<<<<<<<<<<< figure 2 >>>>>>>>>>>>>>>>>>>>>>>>>
The results of the present
study together with the data for AgVP$_2$S$_6$ are shown in
Fig.~\ref{DeJh}~(a). It is remarkable that not only the absolute
values, but also their temperature dependencies are similar for
both materials.
An important observation is that, for both $S=1$
materials, $D_E(T)$ remains rather low in the whole investigated
temperature region. This is in stark contrast to the $S=1/2$
chains where $D_E(T)$ increases by about two orders of magnitude
with decreasing temperature, as is also shown in Fig.~\ref{DeJh}~(a).

The question is whether this is a generic difference, related to
the integrability of the $S=1/2$ and the non-integrability of the
$S=1$ chain model, or is it caused by extrinsic influences. 
To our knowledge, theoretical calculations for the energy
diffusion in $S=1$ HAF chains at $T \lesssim J/k_B$ are absent.
The early analysis by Huber {\em et al.} gives the
high-temperature limiting value for $D^{ht}_E \sim 2 J/\hbar$
\cite{Huber69a},  while a recent numerical study suggests a
higher value of $D^{ht}_E \approx 5.6 J/\hbar$
\cite{Karadamoglou04}. Indeed, at high temperatures $D_E(T)$ of
Y$_2$BaNiO$_5$ is close to this limiting value, see
Fig.~\ref{DeJh}~(a). 
The fact that $D_E(T)$ approaches $D^{ht}_E$ calculated for purely intrinsic interactions \cite{Karadamoglou04},  suggests that intrinsic processes largely determine the energy diffusion in Y$_2$BaNiO$_5$ at high temperatures.

However, the experimental energy diffusion constant for $S=1$ chains does not scale as $T/J$, although such a scaling should hold for intrinsic diffusion. This scaling is also absent for the $S=1/2$ case where predominantly external perturbations, such as spin-phonon interaction
\cite{Shimshoni03},  determine the behavior of spin-mediated heat transport.   
This  suggests that, at least for a part of the excitation spectrum of the $S=1$ chains, 
external perturbations are also important.
Besides phonons, magnetic impurities, that is ions with spin $S \neq 1$, are also
expected to contribute to the scattering of spin excitations. If,
in a certain temperature range, this type of scattering
dominates, one may expect the magnon mean free path  $\ell_s$ to be
close to the average distance between the impurities in the
chains. Information about the concentration of magnetic
impurities can be obtained from the low-temperature increase of
the magnetic susceptibility, if the type of impurities is known
\cite{Payen00,Das04}. For our sample, this is not known, but we
can estimate the lower limit of the average chain segment length
$\ell_{\rm min}^{\rm segment}$ by assuming that all magnetic
impurities have $S=1/2$ and are situated in the chains. A
corresponding fit of the susceptibility data (not shown) of our sample yields
$\ell_{\rm min}^{\rm segment} \sim 60$ ~\AA. In Fig.~\ref{DeJh}~(b) the mean free path of the spin excitations,
estimated via $\ell_s = \kappa_s / (C_s v_s)$ where $v_s$ is the
average velocity of spin excitations, is compared with
$\ell_{\rm min}^{\rm segment}$. 
In our calculations we used $v_s(T)= (Z\hbar)^{-1} \int
\varepsilon^{\prime}_k  f(\varepsilon,T) dk,$ where $Z=\int
f(\varepsilon,T) dk,$  $f(\varepsilon,T) = (\exp (\varepsilon /k_B T) - 1)^{-1}$, and $\varepsilon(k)$ is the dispersion relation. For
magnons in the $S=1$ chain we used the corresponding relations
from the NL$\sigma$M,  $\varepsilon(k) = \left[ V^{2}(ka)^{2} +
\Delta(T)^{2}  \right]^{1/2}$ with $\Delta_0 = 0.41 J$ and $V =
2.49 J$ \cite{Sorensen93}. 

At low temperatures $\ell_s(T)$
approaches $\ell_{\rm min}^{\rm segment}$ suggesting that magnetic
impurities are indeed one of the most important sources of magnon
scattering. For several $S=1/2$ chain compounds a saturation of
$\ell_s(T)$ at low temperatures is observed
\cite{Sologubenko03_Uni}. This is intuitively understood as
resulting from scattering by defects, in analogy with the
boundary scattering in conventional systems. However, after
reaching a peak around 120~K the derived $\ell_s(T)$ of
Y$_2$BaNiO$_5$ decreases again with further decreasing
temperature; a similar behavior was also observed for the $S=1$
compound AgVP$_2$S$_6$ \cite{Sologubenko03_AVPS}. Yet, this
behavior does not contradict the assumption that impurities are
the dominant scatterers at low temperatures, since in gapped 1D
spin systems impurities are expected to give a constant mean free
path only at high temperatures \cite{Orignac03}.  At $T \ll
\Delta$,  the scattering efficiency of point defects is
energy-dependent and the low-energy spin excitations are
scattered by impurities stronger than high-energy excitations
\cite{Orignac03}.  This should lead to a mean free path
decreasing with decreasing temperature for  $T \lesssim \Delta$ in the $S=1$ chains, in contrast to the gapless $S=1/2$ chains, as is illustrated in Fig.~\ref{DeJh}~(c). 
This expectation is in apparent agreement with our experimental results. 

In conclusion, our study of the anisotropic thermal conductivity
of the $S=1$ Haldane chain  system Y$_2$BaNiO$_5$ permits an
evaluation of the contribution of spin excitations to the heat
transport along the chain direction. The corresponding energy
diffusion constant is found to be close to the predicted
high-temperature limiting value. With decreasing temperature, the
spin-related energy diffusion constant of Y$_2$BaNiO$_5$
increases much less than those of $S=1/2$ chain compounds in agreement with theoretical expectations. The
absolute values and the temperature dependence of the magnon mean
free path suggest that magnetic impurities provide the dominating
scattering mechanism for magnons at low temperatures.

We acknowledge useful discussions with J.~A. Mydosh. This work was
supported by the Deutsche Forschungsgemeinschaft through SFB 608


\begin{thebibliography}{10}

\bibitem{Castella95}
H. Castella, X. Zotos, and P. Prelov\v{s}ek, Phys. Rev. Lett.
{\bf 74},  972
  (1995).

\bibitem{Saito96}
K. Saito, S. Takesue, and S. Miyashita, Phys. Rev. E {\bf 54},
2404  (1996).

\bibitem{Zotos96}
X. Zotos and P. Prelov\v{s}ek, Phys. Rev. B {\bf 53},  983
(1996).

\bibitem{Zotos05_Rev}
X. Zotos, J. Phys. Soc. Jpn. {\bf 74},  173   (2005).

\bibitem{Takigawa96_213}
M. Takigawa {\it et~al.}, Phys. Rev. Lett. {\bf 76},  4612
(1996).

\bibitem{Thurber01}
K.~R. Thurber {\it et~al.}, Phys. Rev. Lett. {\bf 87},  247202
(2001).

\bibitem{Sologubenko01}
A.~V. Sologubenko {\it et~al.}, Phys. Rev. B {\bf 64},  054412
(2001).

\bibitem{Sologubenko03_Uni}
A.~V. Sologubenko {\it et~al.}, Europhys. Lett. {\bf 62},  540
(2003).

\bibitem{Haldane83}
F.~D.~M. Haldane, Phys. Lett. {\bf A93},  464  (1983).

\bibitem{Karadamoglou04}
J. Karadamoglou and X. Zotos, Phys. Rev. Lett. {\bf 93},  177203
(2004).

\bibitem{Sologubenko03_AVPS}
A.~V. Sologubenko {\it et~al.}, Phys. Rev. B {\bf 68},  094432
(2003).

\bibitem{darriet93}
J. Darriet and L.~P. Regnault, Solid State Commun. {\bf 86},
409  (1993).

\bibitem{batlogg94}
B. Batlogg, S.-W. Cheong, and L.~W. Rupp~Jr., Physica B {\bf
194-196},  173
  (1994).

\bibitem{yokoo95}
T. Yokoo {\it et~al.}, J. Phys. Soc. Jpn. {\bf 64},  3651  (1995).

\bibitem{sakaguchi96}
T. Sakaguchi {\it et~al.}, J. Phys. Soc. Jpn. {\bf 65},  3025
(1996).

\bibitem{Sulewski94}
P.~E. Sulewski and S.-W. Cheong, Phys. Rev. B {\bf 50},  551
(1994).

\bibitem{Hofmann03}
M. Hofmann {\it et~al.}, Phys. Rev. B {\bf 67},  184502  (2003).

\bibitem{BermanBook}
R. Berman, {\em Thermal conduction in solids} (Clarendon Press,
Oxford, 1976).

\bibitem{Nakamura91}
Y. Nakamura {\it et~al.}, Physica C {\bf 185-189},  1409  (1991).

\bibitem{Hess03}
C. Hess {\it et~al.}, Phys. Rev. Lett. {\bf 90},  197002  (2003).

\bibitem{Sales02}
B.~C. Sales {\it et~al.}, Phys. Rev. Lett. {\bf 88},  095901
(2002).

\bibitem{Yan03}
J.~Q. Yan, J.~S. Zhou, and J.~B. Goodenough, Phys. Rev. B {\bf 68},
104520  (2003).

\bibitem{Jin03}
R. Jin {\it et~al.}, Phys. Rev. Lett. {\bf 91},  146601  (2003).

\bibitem{Berggold06}
K. Berggold {\it et~al.},  Phys. Rev. B {\bf 73},  104430  (2006).


\bibitem{Sologubenko00_lad}
A.~V. Sologubenko {\it et~al.}, Phys. Rev. Lett. {\bf 84},  2714
(2000).

\bibitem{Hess01}
C. Hess {\it et~al.}, Phys. Rev. B {\bf 64},  184305  (2001).

\bibitem{Kudo01}
K. Kudo {\it et~al.}, J. Phys. Soc. Jpn. {\bf 70},  437  (2001).

\bibitem{Ando98}
Y. Ando {\it et~al.}, Phys. Rev. B {\bf 58},  R2913   (1998).

\bibitem{Salce98}
B. Salce {\it et~al.}, Phys. Lett. A {\bf 245},  127   (1998).

\bibitem{Ribeiro05}
P. Ribeiro {\it et~al.}, J. Mag. Magn. Mater. {\bf 290-291},
334  (2005).

\bibitem{Hofmann01}
M. Hofmann {\it et~al.}, Phys. Rev. Lett. {\bf 87},  047202
(2001).

\bibitem{Hess04}
C. Hess and B. B\"uchner, Eur. Phys. J. B {\bf 38},  37  (2004).

\bibitem{Graebner86}
J.~E. Graebner {\it et~al.}, Phys. Rev. B {\bf 34},  5696
  (1986).

\bibitem{Cahill92}
D.~G. Cahill {\it et~al.}, Phys. Rev. B {\bf 46},  6131
  (1992).

\bibitem{Ramirez94}
A.~P. Ramirez {\it et~al.}, Phys. Rev. Lett. {\bf 72},  3108
   (1994).

\bibitem{Jolicoeur94Priv}
T. Jolicoeur and O. Golinelli, Phys. Rev. B {\bf 50}, 9265
(1994); T.~Jolicoeur, private communication.

\bibitem{Huber69a}
D.~L. Huber {\it et~al.}, Phys. Rev. {\bf 186},  534
  (1969).


\bibitem{Sorensen93}
E.~S. Sorensen and I. Affleck, Phys. Rev. Lett. {\bf 71},  1633
(1993).


\bibitem{Shimshoni03}
E. Shimshoni {\it et~al.}, Phys. Rev. B {\bf 68}, 104401  (2003).

\bibitem{Payen00}
C. Payen {\it et~al.}, Phys. Rev. B {\bf 62},  2998  (1990).

\bibitem{Das04}
J. Das {\it et~al.}, Phys. Rev. B {\bf 69},  144404  (2004).

\bibitem{Orignac03}
E. Orignac {\it et~al.}, Phys. Rev. B {\bf 67}, 134426  (2003).

\end{thebibliography}
\end{document}